\documentstyle{l-aa}
\begin{document}
\thesaurus{(03) 02.01.1, 02.16.1, 02.18.5, 11.02.1, 11.10.1, 11.17.3}
\title{Relativistic electron beams in IDV blazars}

\author{Andr\'e R. Crusius-W\"atzel and Harald Lesch}
\offprints{A.R. Crusius-W\"atzel}
\institute{Institut f{\"u}r Astronomie und Astrophysik der Universit{\"at}
M{\"u}nchen, Scheinerstra{\ss}e 1, D-81679 M{\"u}nchen, Germany}

\date{Received date ; accepted date}
\maketitle

\markboth{A.R. Crusius-W\"atzel and H. Lesch:
Relativistic electron beams in IDV blazars}{}

\begin{abstract}
The observed variability of BL Lac objects and Quasars on timescales
$\la 1$ day (intraday variability, IDV)
have revealed radio brightness temperatures up to
$T_{\rm b}\sim 10^{16}-10^{20}$ K. These values challenge the
beaming model with isotropic comoving
radio emission beyond its limits, requiring bulk relativistic motion
with Lorentz factors $\Gamma\ga 100$. We argue in favor of a model
where an anisotropic distribution of relativistic electrons
streams out along the field lines. When this relativistic beam is scattered
in pitch angle and/or hits a magnetic field with components
perpendicular to the beam velocity it starts to emit synchrotron radiation
and redistribute in momentum space.
The propagation of relativistic electrons with Lorentz factor
$\gamma_0\sim 10^2-10^4$ reduces
the intrinsic variability timescale $\Delta t'$ to the observed value
$\Delta t\sim \Delta t'/\gamma_0$
so that the intrinsic brightness temperature is reduced by a factor of order
$\sim 1/\gamma_0^2$, easily below the Inverse Compton limit
of $T_{\rm b}\la 10^{12}$ K. When looking at a single
event we expect the variability time scales $\Delta t$ to be independent of
frequency for a monoenergetic electron beam,
whereas for a beam with a spread out distribution of energies (e.g. power-law)
parallel to the magnetic field the timescales are
shortened towards higher frequencies according to $\Delta t\propto \nu^{-0.5}$.
The observations seem to favor monoenergetic
relativistic electrons which explain several properties of
variable blazar spectra. The production of variable X- and gamma-ray
flux is briefly discussed.


\end{abstract}

\section{Introduction}
It has become a reliable observational fact that active galactic nuclei are
variable on all timescales throughout the whole electromagnetic
spectrum from the radio range up to the gamma-ray-range
(see Wagner and Witzel 1995 for a comprehensive review).

An especially interesting type of variability arises in the radio range on the
timescales of less than one day, called intraday variability (IDV).
IDV at radio frequencies has brought up significant difficulties
with the standard picture of beamed intrinsically isotropic incoherent
synchrotron emission. If the
variability timescale $\Delta t$ is used to calculate the source size by
$l\sim c\Delta t$, where $c$ is the velocity of light the resulting
brightness temperatures reach up to the range $T_{\rm b}=10^{16}-10^{20}$ K,
far in excess of the inverse Compton limit of $T_{\rm b}=10^{12}$ K.
The question arises: What is the physical nature of this very intense
variable radio emission?

We first summarize the properties of the varying sources.
They usually vary on the 5-10\% level at cm wavelengths (Witzel
1992) but it may also be as high as 35\% (Quirrenbach et al. 1992). The
sources that show IDV of the total intensity generally also show IDV of
the polarised flux and the polarisation angle, usually (anti-) correlated.
The typical Blazar that is variable on such short timescales seems to have
the following properties (Wegner 1994):
1.) it reveals a compact structure on VLBI scales (VLA compactness is not sufficient),
2.) associated variability of the polarisation,
3.) if the object is variable at radio frequencies, it is also variable in
the optical,
4.) all BL Lac objects are variable,
5.) sources with a positive spectral index $\alpha_2^{6}\approx 0.2$ between 2 and 6
cm ($I_{\nu}\propto \nu^{\alpha}$) show the largest variability amplitudes.

In the standard picture of beamed intrinsically isotropic emission it is
found that a brightness temperature $T'$ in the comoving frame will
be seen as a {\sl variability} brightness temperature of
$T_{\rm b}\sim D^3T'$, where $D=[\Gamma(1-\beta\cos\theta)]^{-1}$ is the
Doppler factor of the jet (Blandford et al. 1990).
Here $\Gamma=(1-\beta^2)^{-1/2}$ is the bulk Lorentz factor,
$\beta c$ the jet velocity in the observers frame and
$\theta$ the angle between the jet and the line of sight.
The Doppler factors calculated
from the observed superluminal expansion are of the order of $D\la 10$
(Cawthorne 1991). This is
much too low to explain the extreme values of $T_{\rm b}$ found in sources with
IDV which would require $D\ga 100$.
Begelmann et al. (1994) discuss the energetic constraints on highly
relativistic jets ($\Gamma\sim 30-100$ to explain $T_{\rm b}\ga 10^{16}\,{\rm K}$)
and find that the jet must carry kinetic energy fluxes comparable to
the total power of a luminous quasar ($L\ga 10^{47}\, {\rm erg\thinspace s}^{-1}$),
whereas many objects with IDV are low luminosity AGN of the BL Lac type.

A completely different point of view involves coherent emission (Baker
et al. 1988; Krishan and Wiita 1990; Benford 1984, 1992; Lesch and
Pohl 1992).  Coherent radiation easily accounts for huge $T_{\rm b}$, since the
radiation intensity is due to collective emission of all particles within a
coherent volume.  The rather involved physics is well known from plasma
laboratory experiments.  The central argument against coherent radiation
processes in AGN is the limitation of the brightness temperature by induced
Compton scattering and/or Raman scattering (Coppi et al. 1993; Levinson and
Blandford 1995). However, it was shown by Benford and Lesch (1998) that
the saturation arguments are not valid for strong Langmuir turbulence excited by
relativistic electron beams and radiated coherently via inverse Compton scattering.

In this contribution we would like to present an alternative scenario, which
considers the possibility that the energy distribution function of the
synchrotron emitting relativistic
electrons is not isotropic in pitch angle throughout its propagation
in the very center of an AGN. In other words we consider an electron beam, whose
momentum parallel to the magnetic field  is considerably higher than its
momentum perpendicular to the field lines. The advantage of this model is that the synchrotron
brightness temperature of a beam distribution has to be calculated with the square
Lorentz factor of the particles (instead of the bulk Lorentz factor for an
isotropic distribution), which can be of order $10^2-10^4$. Then the
discrepancy of the inverse Compton limit of $10^{12}{\rm K}$ with the observed
values up to $10^{20}{\rm K}$ is resolved.

In the next section we consider the formation and
propagation of relativistic electron beams within the environment of active galactic nuclei.
Here we especially investigate the conditions for isotropization
by magnetohydrodynamic turbulence and how this redistribution towards larger
pitch angles can be avoided in regions with high magnetic field strength.
What follows is a calculation of the resulting
synchrotron radiation of beams for two cases: for a beam with a narrow
energy distribution (quasi-monoenergetic with a clearly defined
low and high energy cutoff) and a beam with a broad energy distribution
(power law). Finally we discuss our results and its implications for
different parts of the electromagnetic spectrum from the radio up to the
gamma-ray range.

\section{Formation and propagation of a relativistic electron beam}

In this section we consider the formation of a beam of relativistic particles
near the central supermassive black hole in an active galactic nucleus
and its propagation in the jet.

\subsection{Acceleration of beam particles}

We consider a spinning back hole with mass $M_{\rm BH}=10^8M_{\sun}M_8$
and angular velocity $\Omega_{\rm BH}$
with an accretion disk that brings in ionized gas and magnetic fields
with strength
$B_0\sim 10^4B_4\ {\rm G}$ near the hole. The rotating magnetic field lines
excert a torque and accelerate particles to high energies
(Gangadhara and Lesch, 1997). The Lorentz factors obtained
in this way are limited by the inverse Compton interaction of the
relativistic particles with the ambient radiation field, basically
UV photons from the disk. This leads to the production of X- and
$\gamma$-rays. The electrons will lose their energy
perpendicular to the magnetic field very rapidly due to synchrotron
radiation in a time
\begin{equation}
t_{\rm S}=5\,10^{-3}\gamma_{0,3}^{-1}B_4^{-2}{\rm \sin}^{-2}\alpha\
{\rm sec}
\end{equation}
where $\alpha$ is the angle between the particle velocity and the
magnetic field and $\gamma_0=10^3\gamma_{0,3}$ is the Lorentz factor
of the relativistic electrons.
The combined action of acceleration and inverse Compton losses
results in a monoenergetic distribution of relativistic electrons
propagating along the magnetic field. The particles accumulate at the energy
where the timescales for gains and losses are equal and balance each other.

Another possibility for the formation of relativistic particle beams
along the magnetic field is by means of reconnecting magnetic fields
in the inner part of the accretion disks and their coronal environments.
Schopper et al. (1998) have
clearly shown that acceleration by magnetic reconnection and energy
losses via synchrotron radiation and/or inverse Compton scattering
leads to beam-like energy distributions. Their
test-particle simulations use realistic magnetohydrodynamical
magnetic field structures originating in sheared, magnetized gas flows
as the back stage for the acceleration process.
Keplerian shear of the field lines leads to the build up of strong magnetic
gradients,
corresponding to high current densities, which finally are dissipated by
localized three-dimensional magnetic reconnection structures with a length $d$.
Three-dimensional magnetic reconnection is equivalent to the presence of
magnetic field aligned electric fields $E_{\parallel}$ along which particles
are efficiently accelerated on very small time scales (Schindler et al. 1991).
Thus, the maximum energy $e E_{\parallel} d$
is reached by almost all particles, only a few particles are influenced
by energy losses. Finally a
quasi-monoenergetic electron distribution function results.

We note that in the central 50 pc and the inner $10^{13}\, {\rm cm}$
of the Milky Way several locations
(Arc and Sgr A$^{*}$) reveal inverted radio spectra, in which the flux increases
with frequency with an exponent 1/3, which is easily interpreted as
optically thin synchrotron radiation of quasi-monoenergetic electron distributions
(Lesch and Reich 1992; Duschl and Lesch 1994). Such a spectrum has also
been observed in M81, which is classified as a ``dead" Seyfert nucleus
(Reuter and Lesch 1996; B\"ottcher et al. 1997).

Obviously central regions of galactic nuclei are preferred physical
environments for the formation of beams. The main reason is
that both mechanisms, centrifugal acceleration and magnetic reconnection are
natural constituents of a rotating magnetized system, including turbulent
gas motions and directed flows (jets), which are all supposed to be necessary
ingredients of models for active galactic nuclei (e.g. Blandford et al. 1990
for a general review)

\subsection{Stability of the beam}

The question arises if these streaming particles can excite Alfv\'en
waves and if they are scattered in pitch angle, which would lead to
isotropisation. It is generally believed that this occurs throughout
the whole jet. Instead we will argue now that this can not be the case
for the innermost part considered here. The phase velocity $v_{\phi}$
of Alfv\'en
waves with frequency $\omega$ and wavenumber $k$ propagating at an angle
$\theta$ with respect to the magnetic field
is given by (Krall and Trivelpiece 1973)
\begin{equation}
v_{\phi}^2={\omega^2\over k^2}={V_{\rm A}^2\ \cos^2\theta}
\left ( {1+V_A^2\over c^2}\right)^{-1}\,.
\end{equation}
Here the parameter called the Alfv\'en velocity is
\begin{equation}
V_{\rm A}=B/\sqrt{4\pi m_{\rm p}n_{\rm p}}
\end{equation}
where $m_{\rm p}$ is the mass and $n_{\rm p}$ is the number density
of the protons  constituting the background plasma
that carries the waves. The numerical value
is given by $V_{\rm A}=2.2\,10^{11}n_{\rm p}^{-1/2}B\ {\rm cm/s}$.
Near the black hole this is larger than the speed of light
for $n_{\rm p}<5.3\ 10^9B_4^2\ {\rm cm}^{-3}$. If one asssumes that
the total kinetic power of the jet is in thermal material this is just
about the maximum average gas density in the jet (Celotti et el., 1998).
Since it is much more likely that the gas is confined in filamentary
structures (probably on the rim of the jet), the density of gas that
can carry Alfv\'en waves in the central regions of the jet is much lower.
In fact from the lack of significant Faraday rotation in BL Lac objects
Celotti et al. (1998) derived $n_{\rm p}f_{\rm V}\la 10^6\,{\rm cm}^{-3}$
at the base of the jet (with radius $R_0=10^{14}\,{\rm cm}$)
where $f_{\rm V}$ is the volume filling factor of the thermal material.
So either the density or the the volume filling factor is low.
They concluded that most of the energy in the jet is carried by the
relativistic particles and the magnetic field in BL Lac objects.

The relativistic electrons can generate and interact with the waves
when the resonance condition
\begin{equation}
\omega - s\Omega - k_{\parallel}v_{\parallel}=0
\end{equation}
is fulfilled. Here $k_{\parallel}=k\cos\theta$ and $v_{\parallel}
=v\cos\alpha$ are the components of the wave vector and the
particle velocity along the magnetic field, respectively, with $\alpha$
being the pitch angle. Further, $s$ is the harmonic number of the
gyromagnetic interaction and $\Omega=\Omega_{\rm e}/\gamma_0$, with
$\Omega_{\rm e}=eB/mc$,
is the relativistic gyrofrequency of the electrons. The case of resonant scattering
of Alfv\'en waves occurs when (Melrose, 1986)
\begin{equation}
|k_{\parallel}v_{\parallel}|\approx |s\Omega |\gg \omega\,.
\end{equation}
From equations (4) and (5) it follows that resonance is only possible when
\begin{equation}
{c \over V_{\rm A}}\left(1+{V_{\rm A}^2\over c^2}\right)^{1/2}\!
\beta\cos\alpha\gg 1
\end{equation}
where $\beta$ is the velocity of the particles normalized by $c$.
In the case $V_{\rm A}\gg c$ this condition reads
$\beta\cos\alpha\gg 1$,
which cannot be fulfilled. When the phase
velocity of the Alfv\'en waves along the field
is close to $c$ the waves that propagate in the medium
are almost purely electromagnetic; these are ineffective in pitch angle scattering.
Achatz and Schlickeiser (1993) find that an electron-positron beam
is completely stable for
\begin{equation}
{\omega_{\rm p}\over\Omega_{\rm e}}<\left({2\over{\gamma_0 -1}}\right)^{1/2}
\end{equation}
where $\omega_{\rm p}=(4\pi n_{\rm e}e^2/m_{\rm e})^{1/2}$ is the plasma
frequency.
This can be compared with the condition as discussed above, since we
can write $V_{\rm A}/c=(m_{\rm e}/m_{\rm p})^{1/2}\Omega_{\rm
e}/\omega_{\rm p}\ga (\gamma_0 m_{\rm e}/m_{\rm p})^{1/2}\sim 1$. So for the
Lorentz factors considered here both are essentially the same.

It seems that in the acceleration region near the
black hole anisotropy is favored and a quasi monoenergetic and
one-dimensional beam of relativistic particles can propagate along the
magnetic field.

\subsection{Disruption of the beam}

The relativistic beam distribution of electrons with number density $N_0$ and
\begin{equation}
N(\gamma)=N_0\ \delta(\gamma-\gamma_0)
\end{equation}
and the background plasma carry a current that produces
a toroidal component $B_{\perp}$ of the magnetic field near the
rim of the jet. The strength scales with the radius $R$ of the jet as
\begin{equation}
B_{\perp}=B_{\perp,0}\left({R\over R_0}\right)^{-1}
\end{equation}
and confines it magnetically.
The poloidal magnetic field in the central part of the jet falls off as
\begin{equation}
B_{\parallel}=B_{\parallel,0}\left({R\over R_0}\right)^{-2}.
\end{equation}
As the jet widens in radius the parallel field ceases
to be important.

It is now interesting to ask when the Alfv\'en velocity $V_{\rm A}$ becomes
less than $c$. The scaling (9) for the toroidal magnetic field
gives a constant value of $V_{\rm A}$ when the particle density
scales according to
\begin{equation}
n_{\rm p}=n_0\left({R\over R_0}\right)^{-2}\,.
\end{equation}
But since the internally dominating poloidal field falls off more
rapidly we find that the Alfv\'en velocity is below the velocity of
light for $B/n_{\rm p}^{1/2}\la 0.1$ or
\begin{equation}
{R\over R_0}\ga 10^2\,B_{0,4}n_{0,6}^{-1/2}
\end{equation}
with $n_0=10^6n_{0,6}\,{\rm cm}^{-3}$.
This means that the beam of relativistic electrons will not be
scattered in pitch angle as long as the jet has not expanded enough.
When condition (12) is fulfilled, i.e. when
$R\ga 10^{16}$ cm and consequently $B_{\perp}\sim 10^2$ G
the beam will rapidly start to
excite Alfv\'en waves which scatter the particles in pitch angle
leading to isotropisation. The flow of the particles becomes unstable
and turbulent as the relativistic beam particles couple to the
background medium. A standing shock will develop some distance L away from
the black hole, which can be identified with the VLBI core. The
relativistic electrons then run into a shock compressed magnetic field
with large perpendicular components and impulsively start to radiate.

\section{Emission from the relativistic electron beam}

The radiation will be
dominated by the perpendicular magnetic field component. The
relativistic particle beam hits the magnetic field with a
Lorentz factor $\gamma_0$ and starts to emit synchrotron radiation.
Although the individual electron will have its emission confined to an
angle of the order of $1/\gamma_0$, the overall emisson of the jet will
be spread over an angle $\varphi\sim 0.1$, the opening angle of the jet.
The spectrum reaches up to the frequency (e.g., Longair 1981)
\begin{equation}
\nu_{\rm c} = {3\over 2}\gamma_0^2\nu_e\sin\alpha
= 4.2\,10^{14}\gamma_{0,3}^2B_2\sin\alpha\ {\rm Hz}
\end{equation}
where $\nu_e=\Omega_{\rm e}/2\pi$ is the nonrelativistic gyrofrequency.
Thus we have emission from radio frequencies up to the optical.
The intensity $I_{\nu}$ is proportional to $\nu^{1/3}$ for frequencies
$\nu\ll\nu_{\rm c}$ and falls off exponentially for $\nu\gg\nu_{\rm c}$.
This naturally gives the observed inverted spectra of compact radio
sources in Blazars. Self-absorption effects may become important at lower
radio-frequencies.

The duration of a radiation event at a fixed location in the
observers frame is given by
\begin{equation}
\Delta t = {\Delta t'\over \gamma_0}\approx {l'\over c}\gamma_0^{-1}
\end{equation}
where $\Delta t'$ is the corresponding time interval in the beam frame
and $l'$ is the intrinsic length scale of the beam.
In the case of $\sim$$10\ {\rm hrs}$ variability
from a $\gamma_0=10^3$ beam the length scale is given by
$l'\sim 10^{18}\ {\rm cm}$.
The brightness temperature that is calculated from light travel time
arguments is then overestimated by a factor $\gamma_0^2$ since
\begin{equation}
T_{\rm B}\propto (c\Delta t)^{-2}=(c\Delta t'/\gamma_0)^{-2}=l'^{-2}\gamma_0^2
\end{equation}
and $l'$ is the comoving size of the emitting region.
The variability timescale is the same
at every frequency, from the optical down to the radio.
The observed variability timescale will be given by (14)
as long as it is larger than the radiative loss time, see
eq. (1).
If the latter becomes larger it smears out the first, because then
the whole beam seems to flare up with the synchrotron life time of
the particles.
From eqs. (1) and (9) we find the condition
\begin{equation}
{R\over R_0}\la 3\,10^3\gamma_{0,3}^{1/2}B_{0,4}\left({\Delta t\over
10\ {\rm hrs}}\right)^{1/2}
\end{equation}
assuming $B_{\perp}$ to be dominant during the emission, according
to what is seen in BL Lac objects.
As long as the radius is limited in such a way, variability is
dominated by Doppler boosting. Indeed jet formation models yield
fairly constant radii over large distances (Camenzind 1996).

If we assume the energy distribution of the beam to be a power-law
(still with zero perpendicular momentum)
\begin{equation}
N(\gamma)\propto \gamma^{-s}
\end{equation}
the emission of synchrotron radiation
at a certain frequency will be dominated by the particles
with a Lorentz factor given by relation (13). The variability timescale
then depends on the frequency. Using (13) and (14) the dependence is found
to be
\begin{equation}
\Delta t\ \propto\ \gamma^{-1}\ \propto\ \nu^{-1/2}\,.
\end{equation}
This means that for a spread in the energy distribution of the radiating
particles the spectral emission will be variable on a shorter timescale at
higher frequencies.

\section{Further evolution of the spectrum}

When the beam starts to dissipate its energy into radiation the
distribution of the particles in energy space will change.
If monoenergetic electrons (8) are injected quasi-continuously and are
confined in the radiating volume for a time long compared to the
synchrotron loss timescale ($\sim 10^{5}$ sec to allow cooling down to
$\gamma\sim 1$ with the parameters considered
here), the particle distribution will become
\begin{equation}
N(\gamma)\propto |\dot\gamma|^{-1}\propto\gamma^{-2}\,.
\end{equation}
This produces a radiation spectrum $I_{\nu}\propto\nu^{-0.5}$ which is
very flat. This is rather stable because the particles now are isotropic
in the comoving jet frame
and do not have the large beaming factor as before but only the one from
the bulk plasma motion, namely $\Gamma$.
Together with the beam radiation one has a highly variable polarised $\nu^{0.3}$
component and a less variable or constant $\nu^{-0.5}$ power-law component with
a different degree and angle of polarisation.
This combination seems to fit to the observations quite well.
A two-component model was already suggested by Qian et al. (1991) to
explain the drastic variations in the linear polarisation.

Assuming that the beam is quasi-steady we also have to consider the
inverse Compton scattering of the power-law component from the
background jet by high energy beam
particles. Since the secondary component radiates isotropically in the
jet frame (also backwards), the beam electrons will have head-on collisions
with those photons (radio to optical). The photons scattered in the
observer's direction will be boosted in frequency by a factor of order
$\gamma_0^2$, thus producing a power-law radiation
component from the optical, over the X-ray to the gamma-ray part of the spectrum.
This emission is beamed towards an observer who also receives the radio to
optical radiation, probably with a time lag.
The timescale for the variability of this high frequency radiation is also
determined by fluctuations in the beam structure and should therefore
be given by eq. (14), i.e., it is of the same order of magnitude as for the
variations in the radio-optical part. In fact this is seen in IDV-blazars
(Wagner and Witzel 1995). Since the variability time-scale depends on the
high Lorentz factor of the beam electrons and not on the bulk Lorentz factor
of the background jet plasma, the constraints on the compactness of the
emitting region are much less stringent.

\section{Summary and discussion}

We consider a model for BL Lac objects and quasars that show variabiliy
on timescales less than a day. The basic ingredient
is a monoenergetic relativistic electron beam.
We discuss the formation and
stability of such an anisotropic distribution and find
that it is stable as long as the magnetic field along the jet is strong
enough. The beam can propagate along the field lines without being
scattered. When the phase velocity of the Alfv\'en waves starts to depart
significantly from the velocity of light the beam will excite
magneto-hydrodynamic waves. The particles will be scattered in pitch angle and
start to emit synchrotron radiation at radio to optical frequencies.
The electrons in the beam are then coupled to the background plasma of
the jet. The spectrum of the monoenergetic particles which are injected
quasi-continuously evolves to a power-law. This gives rise to a second
component in the radiation spectrum. Concerning the high brightness
temperatures derived from variability arguments in IDV blazars,
the upper limit on the compactness of the
emitting region is relaxed by the Lorentz factor of the beam, which can be much
larger than the bulk Lorentz factor of the jet. 

It seems even possible, when the
Lorentz factor of the beam particles is of the order of $\gamma_0\sim 10^6$,
that also highly variable emission in the TeV range observed in several AGN
(e.g. Mkn 421; Gaidos et al. 1996) can be produced this way. There the observed
timescales go down to 0.5 hours. This is beyond the capabilities of models
involving bulk relativistic beaming of intrinsically isotropic distributions
of particles and radiation with particle acceleration at shocks. When the
TeV emission comes from an anisotropic beam distribution the intrinsic length
is only restricted to
$l'\la 10^{20}(\Delta t/1\,{\rm hr})(\gamma_0/10^6)\,{\rm cm}$.
This reduces the opacity of $\gamma-\gamma$ interactions below its
critical value.

For the objects considered here the Lorentz factor of the monoenergetic
electron beam is in the range $10^2-10^4$. This is the basic input
parameter.
The model presented in this paper can account for several features observed
in IDV blazars:

\par\noindent 1.) the high variability brightness temperatures by relativistic
effects
\par\noindent 2.) synchrotron radiation from the monoenergetic beams has an inverted
spectral index as observed in the most variable sources
\par\noindent 3.) the dominant timescale is quasi-stable, which can be understood
in terms of the distance from the black hole where the standing shock developes
\par\noindent 4.) together with the radiation from the reprocessed beam, a
two-component model can be constructed, to explain drastic
changes in the polarisation and its angle
\par\noindent 5.) in such a combination $I_{\nu}\propto a\nu^{0.3} + b\nu^{-0.5}$
one also expects the radio variations to be more pronounced at higher frequencies,
which is seen in the modulation index
\par\noindent 6.) since the synchrotron radiation of the beam reaches
from the radio up to the optical, correlated variability (with the same timescale)
is obviously explained
\par\noindent 7.) a larger variability amplitude at optical frequencies
compared to the radio is expected (and observed)
\par\noindent 8.) correlations of the radio spectral-index with optical flux
(flatter spectrum when optical flux rises) without any measurable time lag
can be understood in terms of the varying $\nu^{1/3}$ emission, which gives
the simultaneous fluctuations at radio and optical frequencies and also a
spectral flattening (increase of spectral index) of the summed components
\par\noindent 9.) because the degree of polarisation of synchrotron radiation
from a monoenergetic distribution depends on the frequency (contrary
to a power-law distribution) this also gives a key to the observed
variations and the frequency dependence of the polarisation and its angle
\par\noindent 10.) if the emission of the two components are polarised (nearly)
orthogonal to each other, this goes along with anticorrelated variations
of the total and the polarised flux (this is the case for BL Lac
objects, where the magnetic field is perpendicular to the jet)
\par\noindent 11.) if the polarisation of the two components is
parallel, correlated variability of the total and polarised flux is expected
(this is the case for quasars, where the magnetic field is oriented parallel
to the jet)
\par\noindent 12.) time lags between total and polarised emission may be attributed
to the reprocession of beam particles into the power-law distribution
\par\noindent 13.) variability at X-ray and gamma-ray
wavelengths (with timescales comparable to the radio/optical)
can be accounted for in head-on collisions
of photons radiated backwards from the power-law component (comoving with the
background jet) off beam electrons by inverse Compton scattering.

This model seems to be compatible with many properties of IDV
blazars. It is now necessary to work out the details of the acceleration
and radiation processes and to compare them with specific objects.

\bigskip
{\sl Acknowledgements.} This work has been supported by the Deutsche
Forschungsgemeinschaft through the grant ME 745/18-1.

{}

\end{document}